\documentclass[prd,a4paper,preprint,preprintnumbers,
nofootinbib,superscriptaddress]{revtex4-2}

\usepackage[a4paper, hdivide={1.91cm,,1.165cm},
vdivide={1.83cm,,3.0cm}]{geometry}

\usepackage[normalem]{ulem}

\usepackage[hyperfootnotes=false]{hyperref}
\hypersetup{linkcolor=red,
citecolor=green,
filecolor=cyan,
urlcolor=magenta}

\usepackage{amsfonts}

\usepackage{amsmath}
\usepackage{amssymb}

\usepackage{bbm}

\usepackage{bbold}

\usepackage{adjustbox}
\usepackage{booktabs}

\usepackage{colortbl}

\usepackage{cancel}

\usepackage{color}
\usepackage{xcolor}

\usepackage{graphicx}
\usepackage{xspace}

\usepackage{units}

\usepackage{slashed}

\usepackage{tensor}

\usepackage{gensymb}

\usepackage{cleveref}

\usepackage{tabularx}
\usepackage{multirow}

\usepackage{setspace}
\usepackage{orcidlink}

\usepackage{caption}
\usepackage{subcaption}

\usepackage{enumerate}
\usepackage{enumitem}

\usepackage[version=3]{mhchem}

\usepackage{lipsum}

\newcommand{{\ia} }{{\i}}
\newcommand{{\Ia} }{{\.I}}

\newcommand{\ra}{\rightarrow}
\newcommand{\la}{\leftarrow}

\def\s2tw{{\rm sin ^2 \theta_{W}}}

\def\beq{\begin{equation}}
\def\eeq{\end{equation}}
\def\bea{\begin{eqnarray}}
\def\eea{\end{eqnarray}}
\def\ve{\vert}
\def\vel{\left|}
\def\ver{\right|}
\def\nnb{\nonumber}

\def\nnb{\nonumber}
\def\la{\langle}
\def\ra{\rangle}

\def\es{ &=& }
\def\ar{&+& }
\def\ek{&-& }
\def\cp{&\times&}

\def\bea{\begin{eqnarray}}
\def\eea{\end{eqnarray}}
\def\beeq{\begin{eqnarray}}
\def\eeeq{\end{eqnarray}}
\def\ve{\vert}
\def\vel{\left|}
\def\ver{\right|}
\def\nnb{\nonumber}

\def\rar{\rightarrow}
\def\lrar{\leftrightarrow}  
\def\nnb{\nonumber}
\def\la{\langle}
\def\ra{\rangle}
\def\ba{\begin{array}}
\def\ea{\end{array}}

\def\xis0{{\Xi^{*0}}}

\def\g5{\gamma_5}

\def\es{\!\!\! &=& \!\!\!}

\def\ar{&+& \!\!\!}
\def\ek{&-& \!\!\!}

\def\cp{&\times& \!\!\!}

\def\mcdot{\!\cdot\!}

\allowdisplaybreaks[1]

\begin{document}


\title{Semileptonic $\Lambda_c \to \Lambda \ell \nu_\ell$ Decays in Light-Cone QCD Sum Rules with $\Lambda_c$ Distribution Amplitudes}

\author{T.~M.~Aliev\,\orcidlink{0000-0001-8400-7370}}
\email{taliev@metu.edu.tr}
\affiliation{Department of Physics, Middle East Technical University, 
Ankara, 06800, Turkey}

\author{S.~Bilmis\,\orcidlink{0000-0002-0830-8873}}
\email{sbilmis@metu.edu.tr}
\affiliation{Department of Physics, Middle East Technical University, 
Ankara, 06800, Turkey}
\affiliation{TUBITAK ULAKBIM, Ankara, 06510, Turkey}

\author{M.~Savci\,\orcidlink{0000-0002-6221-4595}}
\email{savci@metu.edu.tr}
\affiliation{Department of Physics, Middle East Technical University, 
Ankara, 06800, Turkey}

\date{\today}

\def\rar{\rightarrow}
\def\Rar{\Rightarrow}
\def\lrar{\leftrightarrow}
\def\bea{\begin{eqnarray}}
\def\eea{\end{eqnarray}}
\def\ve{\vert}
\def\vel{\left|}
\def\ver{\right|}
\def\nnb{\nonumber}
\def\Rar{\Rightarrow}
\def\nnb{\nonumber}
\def\la{\langle}
\def\ra{\rangle}
\def\ba{\begin{array}}
\def\ea{\end{array}}
\def\es{&=&}
\def\ar{&+&}
\def\ek{&-&}
\def\cp{&\times&}
\def\mcdot{\!\cdot\!}

\begin{abstract}
We study the semileptonic decay of its SU(3) partner, the $\Lambda_c \to \Lambda \ell^+ \nu_\ell$ ($\ell = e, \mu$) transition, within the framework of light-cone QCD sum rules (LCSR) by using the distribution amplitudes of heavy $\Lambda_c$ baryon. The numerical analysis is performed using two different sets of $\Lambda_c$ baryon light-cone distribution amplitudes. The resulting form factors are parametrized by a model-independent $z$-series expansion and used to compute the differential and total decay widths. Our predictions for the branching fractions are in good agreement with the latest BESIII measurements and with lattice-QCD results. 
\end{abstract}

\maketitle

\newpage

\section{Introduction}

\label{sec:intro}

Heavy-flavor baryons provide an important testing ground for our understanding of the interplay between perturbative and nonperturbative aspects of Quantum Chromodynamics (QCD). In particular, semileptonic decays of charmed baryons serve as clean probes of the Standard Model (SM), as the leptonic current factorizes and hadronic uncertainties are confined to a limited set of form factors. Among them, the $\Lambda_c$  baryon occupies a special role as the lightest charmed baryon with a relatively long lifetime and well-measured properties. \(\Lambda_c^+\), undergoes the Cabibbo-favored transition \(c \to s\, \ell^+ \nu_\ell\) predominantly through the channel \(\Lambda_c \to \Lambda \ell \nu_\ell\) (\(\ell = e, \mu\)). Precise determinations of this decay rate depend on the Cabibbo–Kobayashi–Maskawa (CKM) matrix element \(|V_{cs}|\) and on hadronic form factors, providing a stringent internal consistency test of the SM and offering sensitivity to potential new physics contributions.

Significant experimental progress has been achieved in the \(\Lambda_c\) sector in recent years~\cite{Li:2025nzx}.  Earlier theoretical estimates of the branching fraction \(B(\Lambda_c^+ \to \Lambda e^+ \nu_e)\) spanned a wide range, from about \(1.4\%\) to \(9.2\%\)~\cite{Cheng:1995fe,Zhang:2025tki,Meinel:2016dqj,Geng:2022fsr,Gutsche:2015rrt,
Faustov:2016yza,Hussain:2017lir,Zhang:2023nxl,Zhao:2018zcb,
Geng:2020gjh,Li:2021qod,Geng:2019bfz,He:2021qnc}.  The BESIII Collaboration provided the first absolute measurement of the electronic mode in 2015, reporting \(B(\Lambda_c^+ \to \Lambda e^+ \nu_e) = (3.63 \pm 0.38 \pm 0.20)\%\)~\cite{BESIII:2015ysy}. This was soon followed by the first absolute measurement of the muonic channel, $B(\Lambda_c^+ \to \Lambda \mu^+ \nu_\mu) = (3.49 \pm 0.46 \pm 0.27)\%$~\cite{BESIII:2016ffj}. Benefiting from increased luminosity, BESIII subsequently provided the most precise measurement to date for the electronic branching fraction: $B(\Lambda_c^+ \to \Lambda e^+ \nu_e) = (3.56 \pm 0.11 \pm 0.07)\%$~\cite{BESIII:2022ysa}. Notably, recent experiments have progressed beyond determining only the total branching fraction and have begun probing the underlying hadronic dynamics by providing the first direct comparisons of the differential decay rate and form factors with theoretical predictions. A combined analysis of the \(e\) and \(\mu\) channels has further improved the precision of the form factor extraction and enabled the first test of lepton flavor universality in \(\Lambda_c \to \Lambda \ell \nu_\ell\) decays~\cite{BESIII:2023jxv}.

On the theoretical side, the same decay has been studied using various theoretical approaches, including first-principles calculations using Lattice QCD (LQCD)~\cite{Meinel:2016dqj,Zhang:2025tki,Bahtiyar:2021voz}  various quark models (constituent, relativistic, and light-front)~\cite{Gutsche:2015rrt,Faustov:2016yza,Zhao:2018zcb,
Geng:2020gjh,Li:2021qod}, Heavy Quark Effective Theory (HQET)~\cite{Cheng:1995fe}, and QCD sum rules (QCDSR)~\cite{Zhang:2023nxl}.While these methods have provided valuable insight, they often yield mutually inconsistent predictions for the form factors, and their uncertainties are not always fully quantified. In this context, new theoretical studies based on independent techniques and updated inputs are essential to match the increasing experimental precision.

A recent lattice QCD study~\cite{Farrell:2025gis} reported a discrepancy between theoretical predictions and experimental measurements in the semileptonic $\Xi_c \to \Xi \ell \nu_\ell$ decays. Motivated by this tension, in our previous work, we have investigated this decay channel using the light-cone sum rules (LCSR) approach with the distribution amplitudes of the initial $\Xi_c$ baryon~\cite{Aliev:2025zbk}, and our results showed good agreement with the lattice predictions. This naturally raises the question of whether a similar situation could occur in the semileptonic decays of the $\Lambda_c$ baryon, the SU(3) partner of the $\Xi_c$. The present work is devoted to addressing this question.

To compute the necessary form factors, we employ the Light-Cone Sum Rules method. The LCSR technique systematically incorporates nonperturbative effects through an expansion in terms of twist, replacing the short-distance operator product expansion (OPE) with a light-cone expansion.  The reliability of LCSR predictions depends critically on the accuracy of the input distribution amplitudes.

In this study, we analyse the \(\Lambda_c \to \Lambda \ell \nu_\ell\) decay within the LCSR framework by employing  \(\Lambda_c\)-baryon light-cone distribution amplitudes (LCDAs)~\cite{Ball:2008fw,Ali:2012pn}. Our approach is based on the distribution amplitudes of the initial heavy baryon, \(\Lambda_c\), although in principle analogous sum rules can also be formulated using the LCDAs of the final-state light baryon, \(\Lambda\)~\cite{Zhang:2023nxl}. 

The paper is organized as follows. In Section~\ref{sec:formfactors}, we present the theoretical framework and derive the sum rules for the relevant form factors describing the \(\Lambda_c \to \Lambda\) transition.  Section~\ref{sec:numerics} contains the numerical analysis of these form factors, where we also provide results for the branching ratios and compare our findings with those reported in the literature.  Finally, Section~\ref{sec:conclusion} summarizes our conclusions and outlines possible directions for future work.

\section{Light-Cone Sum Rules for the $\Lambda_c \to \Lambda$ Transition}
\label{sec:formfactors}

The effective Hamiltonian responsible for the semileptonic weak decay of the anti-triplet
charmed baryon \(\Lambda_c\) is given by
\bea
  \label{eq4}
\mathcal{H}_{eff} = \frac{G_F}{\sqrt{2}} V_{cs} \, \bar{s} 
\gamma^\mu (1 - \gamma_5) c \, \bar{\nu} \gamma_\mu (1 - \gamma_5) \ell,
\eea
where \(G_F\) is the Fermi coupling constant and \(V_{cs}\) denotes the CKM matrix element.
The weak transition amplitude is obtained by inserting
\(\mathcal{H}_{eff}\) between the initial and final baryon states:
\bea
  \label{eq5}
\mathcal{M}(\Lambda_c \to \Lambda \, \ell \nu_e) =
\frac{G_F}{\sqrt{2}} V_{cs} \, \bar{\nu} \gamma^\mu (1 - \gamma_5) \ell \,
\langle \Lambda (p^\prime, s^\prime) \vert \bar{s} 
\gamma_\mu (1 - \gamma_5) c \vert \Lambda_c(p, s) \rangle.
\eea
The hadronic matrix element is parameterized in terms of six independent form factors as
\bea
  \label{eq6}
\langle \Lambda (p^\prime, s^\prime) \vert \bar{s} \gamma_\mu (1 - \gamma_5) c \vert \Lambda_c(p, s) \rangle 
&=& \bar{u}_{\Lambda}(p^\prime, s^\prime) \Bigg[ \gamma_\mu f_1(q^2) - i \sigma_{\mu\nu}
\frac{q^\nu}{m_{\Lambda_c}} f_2(q^2) + \frac{q_\mu}{m_{\Lambda_c}} f_3(q^2) \nnb \\
&-&  \gamma_\mu \gamma_5 g_1(q^2) + i \sigma_{\mu\nu} \gamma_5 
\frac{q^\nu}{m_{\Lambda_c}} g_2(q^2) - \frac{q_\mu}{m_{\Lambda_c}} 
g_3(q^2) \gamma_5 \Bigg] u_{\Lambda_c}(p, s),
\eea
where \(q = p - p^\prime\) is the momentum transfer, \(m_{\Lambda_c}\) is the mass of the
\(\Lambda_c\) baryon, and \(f_i(q^2)\) and \(g_i(q^2)\) (\(i=1,2,3\))
denote the vector and axial-vector form factors, respectively.

For the calculation of decay widths, it is convenient to introduce helicity amplitudes, which can be expressed directly in terms of the form factors. To express the helicity amplitudes \(H_{\lambda, \lambda_W}^{V(A)}\) in
terms of the form factors \(f_i(q^2)\) and \(g_i(q^2)\), we define
\(\lambda = \pm \tfrac{1}{2}\) as the helicity of the final-state
\(\Lambda\) baryon and \(\lambda_W = t, \pm 1, 0\) as the helicities
of the virtual \(W\) boson.
These amplitudes are obtained from
\bea
  \label{eq7}
H_{\lambda, \lambda_W} &=& \langle \Lambda(p^\prime, \lambda) \vert \bar{s} \gamma_\mu 
(1 - \gamma_5) c \vert \Lambda_c(p, \lambda) \rangle \,
\varepsilon^{*\mu}(\lambda_W)~.
\eea
Following standard procedures (see, e.g.,
\cite{Aliev:2025zbk,Gutsche:2015rrt}),
we obtain the explicit expressions:
\bea
  \label{eq9}
H_{+1/2, t}^V &=& \sqrt{\frac{Q_+}{q^2}} \left[ m_- f_1 + \frac{q^2}{m_{\Lambda_c}} f_3
\right], \nnb \\
H_{+1/2, +1}^V &=& \sqrt{2Q_-} \left[ f_1 + \frac{m_+}{m_{\Lambda_c}} f_2 \right], \nnb \\
H_{+1/2, 0}^V  &=& \sqrt{\frac{Q_-}{q^2}} \left[ m_+ f_1 + \frac{q^2}{m_{\Lambda_c}} f_2 \right], \nnb \\
H_{+1/2, t}^A &=& \sqrt{\frac{Q_-}{q^2}} \left[ g_1 m_+ - \frac{q^2}{m_{\Lambda_c}} g_3 \right], \nnb \\
H_{+1/2, +1}^A &=& \sqrt{2Q_+} \left[ g_1 - \frac{m_-}{m_{\Lambda_c}} g_2 \right], \nnb \\
H_{+1/2, 0}^A  &=& \sqrt{\frac{Q_+}{q^2}} \left[ g_1 m_- - \frac{q^2}{m_{\Lambda_c}} g_2
\right],
\eea
where
\bea
m_\pm &=& m_{\Lambda_c} \pm m_\Lambda, \nnb \\
Q_\pm &=& m_\pm^2 - q^2, \nnb
\eea
and \(m_{\Lambda_c}\) and \(m_\Lambda\) are the masses of the \(\Lambda_c\) and \(\Lambda\) baryons,
respectively.

The helicity amplitudes satisfy the following symmetry relations,
\bea
H^V_{-\lambda, -\lambda_W} &=& - H^V_{\lambda, \lambda_W}, \nnb \\
H^A_{-\lambda, -\lambda_W} &=& + H^A_{\lambda, \lambda_W}. \nnb
\eea

Using these helicity amplitudes, the differential decay width can be obtained as
\bea
\label{eq10}
\frac{d\Gamma}{dq^2} = \frac{G_F^2 \lvert V_{cs}\rvert^2}{192 \pi^3} \frac{ (q^2 - m_\ell^2)^2
\lvert\vec{p}\rvert}{m_{\Lambda_c}^2 q^2} H_{tot},
\eea
where
\bea
\label{eqh}
H_{tot} &=& \left(1 + \frac{m_\ell^2}{2q^2}\right) \left[ \lvert H_{+1/2, +1}\rvert^2 + 
\lvert H_{-1/2, -1}\rvert^2 + \lvert H_{+1/2, 0}\rvert^2 + \lvert H_{-1/2, 0}\rvert^2 \right] \nnb \\
&+& \frac{3m_\ell^2}{2q^2} \left[ \lvert H_{+1/2, t}\rvert^2 + \lvert H_{-1/2, t}\rvert^2 \right]. 
\eea
The magnitude of the three-momentum of the final-state baryon is
\bea
\lvert \vec{p}\rvert  = \frac{1}{2m_{\Lambda_c}} \lambda^{1/2}(m_{\Lambda_c}^2,
m_\Lambda^2, q^2), \nnb
\eea
where the Källén function $\lambda$ is defined as
\bea
\lambda(a, b, c) = a^2 + b^2 + c^2 - 2ab - 2ac - 2bc. \nnb
\eea
In Eq.~\eqref{eqh} $H_{\lambda, \lambda_w}$ is defined as
\bea
H_{\lambda, \lambda_W} = H^V_{\lambda, \lambda_W} - H^A_{\lambda, \lambda_W}.
\eea
To determine the form factors responsible for the \(\Lambda_c \to \Lambda\)
transition, we use the light-cone QCD sum rule method. Within this
framework, we start from the correlation function
\bea
\Pi_\mu = i \int d^4x \, e^{ip^\prime  x} 
\langle 0 \vert  \eta_\Lambda(x) \, \bar{J}_\mu(0) \vert \Lambda_c(p) \rangle, \nnb
\eea
where \(\eta_\Lambda(x)\) is the interpolating current for the \(\Lambda\) baryon,
defined as
\bea
\eta_\Lambda(x) &=& \frac{2}{\sqrt{6}} \epsilon^{abc} \sum_{\ell=1}^{2} \Big\{
2 \big[ u^a(x) A^\ell d^b(x) \big] B^\ell s^c(x) 
+ \big[ u^a(x) A^\ell s^b(x) \big] B^\ell d^c(x) \nnb \\
&+& \big[ s^a(x) A^\ell d^b(x) \big] B^\ell u^c(x)
\Big\}, \nnb
\eea
and \(\bar{J}_\mu = \bar{s} \gamma_\mu (1 - \gamma_5) c\) denoting the weak transition current.
Here \(a, b, c\) are color indices, \(A^1 = \mathbb{1}\), \(A^2 = C \gamma_5\),
\(B^1 = \gamma_5\), \(B^2 = \beta \mathbb{1}\), and \(C\) is the charge conjugation operator.
The interpolating current \(\eta_\Lambda\) is constructed to match the quantum numbers
of the \(\Lambda\) baryon and is used in the evaluation of the QCD side of the correlation function.

Taking into account the Dirac equation \((\slashed{p} - m) u(p) = 0\),
the correlation function can be expressed in terms of independent invariant amplitudes.
For the vector and axial-vector currents, six invariant amplitudes arise in each case.
The full decomposition of the correlation function reads
\bea
\label{eq11}
\Pi_\mu &=& \big[  \Pi_1 \, v_\mu \slashed{q} + \Pi_2 \, \gamma_\mu  \slashed{q} + 
 \Pi_3 \, q_\mu + \Pi_4 v_\mu + \Pi_5 \gamma_\mu + \Pi_6 \, q_\mu \slashed{q} \big] u(v) \nnb \\
 &-& \big[\Pi_7 \, v_\mu \slashed{q} + \Pi_8 \, \gamma_\mu  \slashed{q} + 
 \Pi_9 \, q_\mu + \Pi_{10} v_\mu + \Pi_{11} \gamma_\mu + \Pi_{12} \, q_\mu \slashed{q}  \big] \gamma_5 \, u(v),
\eea
where we introduced \(p_\mu = m_{\Lambda_c} v_\mu\) and used the heavy-baryon
velocity notation,
\bea
\vert \Lambda_c(p)\rangle = \sqrt{m_{\Lambda_c}} \, \vert \Lambda_c(v)\rangle. \nnb
\eea

Using the definition \(\langle 0 \vert \eta_\Lambda \vert \Lambda(p^\prime) \rangle = \lambda \, u(p^\prime)\),
together with the decomposition of the \(\Lambda_c \to \Lambda\) transition form factors
and after performing the spin summation over the \(\Lambda\) baryon,
one obtains the hadronic representation of the correlation function in terms of
invariant amplitudes.

\bea
\label{eq12}
\Pi_\mu(p, q) &=& \frac{\lambda}{m_\Lambda^{2} - p^{\prime 2}} \Bigg\{
f_1 \bigg[ 2 m_{\Lambda_c} v_\mu + 
(m_\Lambda - m_{\Lambda_c})\gamma_\mu - 2q_\mu +
\gamma_\mu \slashed{q} \bigg] \nnb \\
&+& \frac{f_2}{m_{\Lambda_c}} \Big[ 2 m_{\Lambda_c} v_\mu \slashed{q} + 
(m_\Lambda^2 - m_{\Lambda_c}^2)\gamma_\mu +
 (m_\Lambda + m_{\Lambda_c})\gamma_\mu \slashed{q} - 
q_\mu (m_\Lambda + m_{\Lambda_c}) - q_\mu \slashed{q} \Big] \nnb \\
&+& \frac{f_3}{m_{\Lambda_c}} \, q_\mu ( m_\Lambda + m_{\Lambda_c} - \slashed{q} ) \nnb \\
&-& g_1 \bigg[ 2 m_{\Lambda_c} v_\mu + (m_\Lambda + m_{\Lambda_c})\gamma_\mu - 2q_\mu + 
\gamma_\mu \slashed{q} \bigg] \gamma_5 \nnb \\
&-& \frac{g_2}{m_{\Lambda_c}} \Big[ 2 m_{\Lambda_c} v_\mu \slashed{q} + (m_\Lambda^2 - 
m_{\Lambda_c}^2)\gamma_\mu + (m_\Lambda - m_{\Lambda_c}) \gamma_\mu \slashed{q}  - 
q_\mu (m_\Lambda - m_{\Lambda_c}) - q_\mu \slashed{q} \Big] \gamma_5 \nnb \\
&-& \frac{g_3}{m_{\Lambda_c}} \, q_\mu ( m_\Lambda - m_{\Lambda_c} - \slashed{q} ) \gamma_5
\Bigg\} u(v).
\eea
This hadronic representation will subsequently be matched to the QCD calculation in the next step.

We now turn our attention to the calculation of the correlation function from QCD side. The external four-momenta are taken space-like, \(p^{\prime 2}, q^2 \ll m_c^2\), so that the
correlation function can be expanded near the light cone (\(x^2 \simeq 0\)).
In this regime, the operator product expansion (OPE) is obtained in terms
of the light-cone distribution amplitudes of the \(\Lambda_c\) baryon, ordered by twist,
convoluted with the hard-scattering amplitudes involving the virtual \(s\)-quark propagator.

After applying Wick’s theorem, the correlation function takes the form
\bea
  \label{eq13}
    (\Pi_\mu)_\rho &=&  \frac{2}{\sqrt{6}} \epsilon^{abc} \sum_{\ell=1}^2 \int d^4 x \, e^{i p^\prime x}
(A^\ell)_{\alpha \beta} (B^\ell)_{\rho \gamma} (\gamma_\mu (1-\gamma_5))_{h \eta} \nnb \\
&\times& \Big\{ 2 S_{\gamma h} \langle 0 \vert u_\alpha^a(x) d_{\beta}^b(x) c_\eta^c(0) \vert \Lambda_c \rangle
+ S_{\beta h} \langle 0 \vert u_\alpha^a(x) d_\gamma^b(x) c_\eta^c(0) \vert \Lambda_c \rangle \nnb \\
&+& S_{\alpha h} \langle 0 \vert u_\gamma^a(x) d_\beta^b(x) c_\eta^c(0) \vert \Lambda_c \rangle 
\Big\},
\eea
where \(S\) denotes the s-quark propagator.

The matrix element
\bea
\langle 0 \vert u^a_\alpha(x) \, d^b_\beta(x) \, c_\gamma^c(0) \vert \Lambda_c(v) \rangle, \nnb
\eea
is parameterized in terms of the light-cone distribution amplitudes of the
\(\Lambda_c\) baryon~\cite{Ball:2008fw}. It can be written as
\bea
\label{eq14}
\epsilon^{abc} \langle 0 \vert u^a_\alpha(t_1 n) \, d^b_\beta(t_2 n) \,
h_\gamma^c(0) \vert \Lambda_c(v) \rangle 
= \sum_{i=1}^{4} \alpha^\prime_i \, [\Gamma_i]_{\beta \alpha} \, (u_i(v))_\gamma,
\eea
where \(h_c^\gamma\) is the heavy-quark field in HQET, \(\Gamma_i\) denote the relevant Dirac structures,
\(u_\gamma(v)\) is the heavy baryon spinor, \(\alpha^\prime_i\) are coefficient functions
defined below, and \(\psi_i\) are the light-cone DAs of the \(\Lambda_c\) baryon.

The explicit expressions of the coefficients are
\bea
\label{eq15}
\alpha^\prime_1 &=& \tfrac{1}{8} f^{(2)} \, \psi_2(t_1, t_2), \nnb \\
\alpha^\prime_2 &=& \tfrac{1}{8} f^{(1)} \, \psi_3^\sigma(t_1, t_2), \nnb \\
\alpha^\prime_3 &=& \tfrac{1}{4} f^{(1)} \, \psi_3^s(t_1, t_2), \nnb \\
\alpha^\prime_4 &=& \tfrac{1}{8} f^{(2)} \, \psi_4(t_1, t_2),
\eea
with the corresponding Dirac structures
\bea
\label{eq16}
\Gamma_1 &=& \slashed{\bar{n}} \gamma_5 C^{-1}, \nnb \\
\Gamma_2 &=& i \sigma_{\mu\nu} \, \bar{n}^\mu n^\nu \gamma_5 C^{-1}, \nnb \\
\Gamma_3 &=& \gamma_5 C^{-1}, \nnb \\
\Gamma_4 &=& \slashed{n} \gamma_5 C^{-1}.
\eea

Here, the subscript \(i\) in \(\psi_i\) indicates the twist of the DA.
The light-cone vectors are defined as
\bea
n_\mu = \frac{x_\mu}{v  x}, \qquad
\bar{n}_\mu = 2 v_\mu - n_\mu. \nnb
\eea

The distribution amplitudes \(\psi(t_1, t_2)\) in coordinate space
are related to their momentum-space representation \(\psi(u, \omega)\) through the
Fourier transform
\bea
\psi(t_1, t_2) = \int_0^1 du \int_0^\infty d\omega \, \omega \,
e^{-i\omega(t_1 \bar{u} + t_2 u)} \, \psi(u, \omega), \nnb
\eea
where \(\omega\) denotes the total momentum carried by the two light 
quarks and \(\bar{u} = 1 - u\).
In the special case \(t_1 = t_2 = v  x\), this simplifies to
\bea
\psi(t_1, t_2) = \int_0^1 du \int_0^\infty d\omega \, \omega \, 
e^{-i\omega v  x} \, \psi(u, \omega).\nnb 
\eea

Using these definitions, the QCD representation of the correlation function can be written as
\bea
\label{eq17}
(\Pi_\mu^{\text{QCD}})_\rho &=& \frac{2}{\sqrt{6}} \, i \int d^4x 
\int_0^1 du \int_0^\infty d\omega \, 
\omega \, e^{i(p^\prime - \omega v)   x} 
\sum_{\ell = 1}^{2} \sum_{i = 1}^{4}  \alpha_i \nnb \\
&\times& (A^\ell)_{\alpha\beta} (B^\ell)_{\rho\gamma} 
(\gamma_\mu (1-\gamma_5))_{h \eta} \nnb \\
&\times& \Big\{
2 (\Gamma_i)_{\beta \alpha} S_{\gamma h}(x)
- (\Gamma_i)_{\gamma \alpha} S_{\beta h}(x)
+ (\Gamma_i)_{\alpha \gamma} S_{\beta h}(x)
\Big\} u_\eta(v)~.
\eea

After performing the Fourier transformation of the propagator and integrating
over \(x\), the final expression for the QCD side of the correlation function becomes
\bea
\label{eq18}
\Pi_\mu^{\text{QCD}} &=& \frac{2}{\sqrt{6}} \int_0^1 du \int_0^\infty 
\omega \, d\omega \sum_{\ell=1}^{2} \sum_{i=1}^{4} \alpha_i \nnb \\
&\times& \Bigg\{
2 \, \text{Tr}[\Gamma_i A^\ell] \, B^\ell \, 
\frac{\slashed{p}^\prime - \omega \slashed{v} + m_s}{m_s^2 - (p^\prime - \omega v)^2} \, 
\gamma_\mu (1 - \gamma_5) \nnb \\
&+& B^{\ell} \Gamma_i A^\ell \, 
\frac{\slashed{p}^\prime - \omega \slashed{v} + m_s}{m_s^2 - (p^\prime - \omega v)^2} \, 
\gamma_\mu (1 - \gamma_5) \nnb \\
&+& B^\ell \, \Gamma_i^T A^{\ell T} \, 
\frac{\slashed{p}^\prime - \omega \slashed{v} + m_s}{m_s^2 - (p^\prime - \omega v)^2} \, 
\gamma_\mu (1 - \gamma_5)
\Bigg\} u(v).
\eea

The final step in obtaining the sum rules for the \(\Lambda_c \to \Lambda\)
transition form factors is to identify the same Lorentz structures from both 
the hadronic and QCD representations of the correlation function and match 
their coefficients.

By isolating the structures proportional to 
\( \slashed{q} v_\mu  \),  \(\gamma_\mu \slashed{q}\), \(q_\mu \), \( \slashed{q} \gamma_5 v_\mu \),  \( \gamma_\mu \slashed{q} \gamma_5 \), and \( \gamma_5 q_\mu\) , we arrive at the following sum 
rules for the form factors, respectively:
\bea
\label{eq19}
   \frac{2 \lambda_\Lambda }{m_{\Lambda}^2 - p^{\prime 2}} f_2 \es \Pi_1, \nnb \\
   \frac{\lambda_\Lambda}{m_{\Lambda}^2 - p^{\prime 2}} \left[ f_1 +
   \frac{m_\Lambda + m_{\Lambda_c}}{m_{\Lambda_c}} f_2 \right] \es \Pi_2, \nnb \\
- \frac{\lambda_\Lambda}{m_{\Lambda}^2 - p^{\prime 2}} \left[ 2 f_1 +
  \frac{m_\Lambda + m_{\Lambda_c}}{m_{\Lambda_c}} f_2 -
  \frac{m_\Lambda + m_{\Lambda_c}}{m_{\Lambda_c}} f_3 \right] \es \Pi_3, \nnb \\
 - \frac{2 \lambda_\Lambda }{m_{\Lambda}^2 - p^{\prime 2}} g_2 \es \Pi_4, \nnb \\
 - \frac{\lambda_\Lambda}{m_{\Lambda}^2 - p^{\prime 2}} \left[ g_1 +
   \frac{m_\Lambda - m_{\Lambda_c}}{m_{\Lambda_c}} g_2 \right] \es \Pi_5, \nnb \\
  \frac{\lambda_\Lambda}{m_{\Lambda}^2 - p^{\prime 2}} \left[ 2 g_1 +
  \frac{m_\Lambda - m_{\Lambda_c}}{m_{\Lambda_c}} g_2 -
  \frac{m_\Lambda - m_{\Lambda_c}}{m_{\Lambda_c}} g_3 \right] \es \Pi_6.
\eea

The invariant functions \(\Pi_i\) can, in general, be represented as
\bea
\label{eq20}
\Pi_i = \int_0^1 du \int \sigma d\sigma \left[
\frac{\rho_i^{(1)}(\sigma)}{\bar{\sigma} \Delta} + 
\frac{\rho_i^{(2)}(\sigma)}{\bar{\sigma}^2 \Delta^2} + 
\frac{\rho_i^{(3)}(\sigma)}{\bar{\sigma}^3 \Delta^3}
\right],
\eea
where
\bea
\Delta &=& p^{\prime 2} - s(\omega), \nnb \\
s(\omega) &=& \frac{m_s^2 + m_{\Lambda_c}^2 \bar{\sigma} - 
\sigma q^2}{\bar{\sigma}}, \nnb \\
\sigma &=& \frac{\omega}{m_{\Lambda_c}}, \qquad \bar{\sigma} = 1 - \sigma. \nnb 
\eea
Explicit expressions for the spectral densities \(\rho_i^{(j)}\) are 
lengthy and are therefore omitted here. To suppress contributions from higher states and the continuum,
we invoke quark–hadron duality in the \(\Lambda\) channel, replacing 
the hadronic spectral density with the integral of the OPE spectral density 
above an effective threshold \(s_0\). Applying a Borel transformation in 
\(p^{\prime 2}\), the resulting relation is
\bea
\int_0^{s_0} ds\, \frac{I_n}{(p^{\prime 2} - s)^n}
&=& \int_0^{\sigma_0} d\sigma\, (-1)^n \frac{e^{-s(\sigma)/M^2} 
I_n}{(n-1)! (M^2)^{n-1}} \nnb \\
&+& \frac{(-1)^n}{(n-1)!} e^{-s/M^2} \sum_{j=1}^{n-1} 
\frac{1}{(M^2)^{n-j-1}} \frac{1}{s^\prime}
\left( \frac{d}{d\sigma} \frac{1}{s^\prime} \right)^{j-1} I_n,\nnb 
\eea
where \(M\) is the Borel parameter, and
\bea
s^\prime &=& \frac{ds}{d\sigma}, \nnb \\
I_n(\sigma) &=& \frac{\sigma \rho_n}{\bar{\sigma}}, \nnb \\
\left( \frac{d}{d\sigma} \frac{1}{s^\prime} \right)^{j-1} I_n(\sigma) 
&\Rightarrow& \left[ \frac{d}{d\sigma} \frac{1}{s^\prime} \left( \frac{d}{d\sigma} 
\frac{1}{s^\prime} \cdots \right) \right] I_n(\sigma), \nnb \\
\sigma_0 &=& \frac{s + m_\Lambda^2 - q^2 - \sqrt{4(m_s^2 - s_0)m_\Lambda^2 + 
(m_\Lambda^2 + s_0 - q^2)^2}}{2m_\Lambda^2}. \nnb 
\eea

As a result, the sum rules for the form factors can be expressed in the compact form
\bea
F_i^{\text{form factor}} = \frac{1}{K_i} e^{m_p\Lambda^2 / M^2} 
\sum_{n=1}^{\infty} \mathcal{B}_i^{(n)}, \nnb
\eea
where \(F_i \in \{ f_i, g_i \}\) denote the form factors,
\(K_i\) are kinematic factors containing residues, 
masses, and normalization constants,
\(i = 1, \dots, 6\) labels the corresponding form factor,
and \(\mathcal{B}_i^{(n)}\) represents the Borel-transformed and subtracted 
invariant amplitude.

\section{Numerical Analysis}
\label{sec:numerics}

In this section, we present the numerical analysis of the sum rules for the 
\(\Lambda_c \to \Lambda\) transition form factors.
As discussed earlier, the main nonperturbative inputs of the light-cone sum rules
are the distribution amplitudes.

In this work, we adapt two different sets of light-cone distribution amplitudes (LCDAs) of the $\Lambda_b$ baryon obtained in ~\cite{Ball:2008fw,Ali:2012pn} and apply the same functional forms to the $\Lambda_c$ by replacing the heavy-quark parameters accordingly. This approach assumes that the shape of the heavy-baryon LCDAs is largely insensitive to the heavy-quark flavor, while possible $1/m_c$ corrections are neglected. The explicit expressions of the DAs are:
The explicit expressions of first set of DAs (named as DAs-set I) obtained in\cite{Ball:2008fw} 
\bea
  \label{eq22}
    \psi_{2}(\omega, u) &=& \frac{15}{2} A^{-1} \omega^{2} \bar{u} 
u \int_{\omega / 2}^{s_{0}} d s \, e^{-s / y}(s-\omega / 2), \nnb \\
    \psi_{4}(\omega, u) &=& 5 A^{-1} 
\int_{\omega / 2}^{s_{0}} d s \, e^{-s / y}(s-\omega / 2)^{3}, \nnb \\
    \psi_{3 s}(\omega, u) &=& \frac{15}{4} A^{-1} \omega 
\int_{\omega / 2}^{s_{0}} d s \, e^{-s / y}(s-\omega / 2)^{2}, \nnb \\
\psi_{3 \sigma}(\omega, u) &=& \frac{15}{4} A^{-1} \omega(2 u-1) 
\int_{\omega / 2}^{s_{0}} d s \, e^{-s / y}(s-\omega / 2)^{2},
\eea
where
\bea
  A=\int_{0}^{s_{0}} d s \, s^{5} e^{-s / y}. \nnb
\eea
Here, \(y\) and \(s_{0}\) denote the Borel parameter and the continuum threshold,
respectively~\cite{Ball:2008fw}.
From the QCD sum-rule analysis, \(y\) is found to vary in the range
\(0.4~\mathrm{GeV} < y < 0.8~\mathrm{GeV}\), while the continuum threshold
is fixed at \(s_{0} = 1.2~\mathrm{GeV}\).

The second set of \(\Lambda_c\) DAs was obtained in Ref.~\cite{Ali:2012pn} and takes the form
\bea
\label{eq21}
\psi_2(u, \omega) &=& \omega^2 u \bar{u} \sum_{i=0}^{2} \frac{a_i}{\varepsilon_i^4} \,
\frac{C_i^{3/2}(2u - 1)}{|C_i^{3/2}|^2} \, e^{-\omega/\varepsilon_i}, \nnb \\
\psi_3^{(\sigma, s)}(u, \omega) &=& \frac{\omega}{2} \sum_{i=0}^{2} 
\frac{a_i}{\varepsilon_i^2} \, \frac{C_i^{1/2}(2u - 1)}{|C_i^{3/2}|^2} \, 
e^{-\omega/\varepsilon_i}, \nnb \\
\psi_4(u, \omega) &=& \sum_{i=0}^{2} \frac{a_i}{\varepsilon_i^2} \, 
\frac{C_i^{1/2}(2u - 1)}{|C_i^{1/2}|^2} \, e^{-\omega/\varepsilon_i},
\eea
where \(C_i^\ell(2u - 1)\) are Gegenbauer polynomials, and the
normalization factor is defined as
\bea
|C_i^\ell|^2 = \int_0^1 du \, \big[ C_i^\ell(2u - 1) \big]^2. \nnb
\eea
The values of the parameters $a_i$ and $\varepsilon_i$ are taken from Ref.~\cite{Ali:2012pn} with the parameter $A$ fixed to $0.5$.

The numerical values of the input parameters entering the sum rules are listed in
Table~\ref{tab:input-params}.

\setlength{\tabcolsep}{8pt} 
\begin{table}[ht]
\centering
\caption{Input parameters used in the analysis.}
\label{tab:input-params}
 \begin{tabular}{lr}
\toprule
Parameter & Value \\ 
\midrule
$m_c(\overline{m_c})$ & $1.273 \pm 0.046$ GeV~\cite{ParticleDataGroup:2024cfk} \\
$m_s(2\,\text{GeV})$ & $93.5 \pm 0.8$ MeV~\cite{ParticleDataGroup:2024cfk} \\
$f^{(1)} = f^{(2)}$ & $(2.3 \pm 0.2) \times 10^{-2}$ GeV$^3$~\cite{Groote:1996em} \\
$m_{\Lambda_c}$ & $2.286$ GeV~\cite{ParticleDataGroup:2024cfk} \\
$m_\Lambda$ & $1.115$ GeV~\cite{ParticleDataGroup:2024cfk} \\
\bottomrule
\end{tabular}
\end{table}

The sum rules involve three auxiliary parameters: the Borel mass \(M^2\),
the continuum threshold \(s_0\), and the mixing parameter \(\beta\) (here we take $\beta = -1$, i.e., Ioffe current) that appears
in the interpolating current.
Physical observables should not depend on these parameters, so our first task
is to determine their working regions.

The continuum threshold \(s_0\) is chosen such that the two-point sum rules
reproduce the experimentally measured \(\Lambda\) baryon mass within
approximately 10\% accuracy.
Our analysis indicates that this requirement is satisfied for
\(2.0~\text{GeV}^2 \le s_0 \le 4.0~\text{GeV}^2\).
Similarly, the allowed range of the Borel parameter is determined by ensuring
that higher-twist and continuum effects remain subdominant compared to the
leading-twist contributions.
This condition is met for
\(1.0~\text{GeV}^2 \le M^2 \le 3.0~\text{GeV}^2\).

With these auxiliary parameters fixed, we next investigate the
\(q^2\)-dependence of the form factors.
The physical range of momentum transfer for the semileptonic decay
\(\Lambda_c \to \Lambda \ell \nu\) is
\begin{equation}
m_\ell^2 < q^2 < (m_{\Lambda_c} - m_{\Lambda})^2, \nnb 
\end{equation}
but the LCSR predictions are most reliable near \(q^2 \simeq 0\),
and more precisely in the region \(q^2 \lesssim 0.5~\text{GeV}^2\).
To extend the form factors across the entire physical region, we adopt the
\(z\)-series expansion~\cite{Bourrely:2008za},
\begin{equation}
f(q^2) = \frac{1}{1 - q^2/m_{\text{pole}}^2}
\left\{ a_0^f + a_1^f\, z(q^2) + a_2^f\, z^2(q^2) \right\}, \nnb 
\end{equation}
where the conformal variable \(z\) is defined as
\begin{equation}
z(q^2) = \frac{\sqrt{t_+ - q^2} - \sqrt{t_+ - t_0}}{\sqrt{t_+ - q^2} + 
\sqrt{t_+ - t_0}}, \nnb 
\end{equation}
with \(t_+ = (m_{\Lambda_c} + m_{\Lambda})^2\).
Here \(m_{\text{pole}}\) corresponds to the lowest-lying resonance with the same
quantum numbers as the current mediating the transition:
\begin{align*}
m_{\text{pole}} = 
\begin{cases}
2.010~\text{GeV} & \text{for } f_1, f_2, \\
2.413~\text{GeV} & \text{for } g_1, g_2, \\
2.300~\text{GeV} & \text{for } f_3, \\
1.870~\text{GeV} & \text{for } g_3. \nnb
\end{cases}
\end{align*}

The fitting parameters $a_0^f$, $a_1^f$, and $a_2^f$ for each
form factor are determined by matching the $z$-series parametrization to the
LCSR predictions in the region $q^2 < 0.5~\text{GeV}^2$.
To simultaneously account for uncertainties in the input parameters and form
factors, we employ a Monte Carlo analysis. In particular, we generate 5000
samples by randomly varying the input parameters within their quoted
uncertainties and compute the corresponding form factors at $q^2=0$. 
The resulting distributions are used to extract the central values and uncertainties of the form factors, which are summarized in Table~\ref{tab:formfactors}. For illustration, only the results obtained using the DA set~I are shown in Fig.~1.
\begin{table}[htb]
  \centering
  \caption{Form factors for the $\Lambda_c \to \Lambda$ transition at
  $q^2=0$, obtained using two different sets of $\Lambda_c$ DAs.}
  \label{tab:formfactors}
  \begin{tabular}{ccc}
    \toprule
    Form Factor & Set I~\cite{Ball:2008fw} & ~~Set II~\cite{Ali:2012pn} \\
    \midrule
    $f_1(0)$ & $-0.32 \pm 0.13$~~&~~$-0.65 \pm 0.08$ \\
    $f_2(0)$ & $0.21 \pm 0.09$~~&~~$0.44 \pm 0.05$ \\
    $f_3(0)$ & $-1.52 \pm 0.41$~~&~~$-2.37 \pm 0.20$ \\
    $g_1(0)$ & $0.65 \pm 0.11$~~&~~$0.79 \pm 0.04$ \\
    $g_2(0)$ & $0.22 \pm 0.09$~~&~~$0.44 \pm 0.05$ \\
    $g_3(0)$ & $-0.22 \pm 0.09$~~&~~$-0.44 \pm 0.05$\\
    \bottomrule
  \end{tabular}
\end{table}
Using the form factors listed in Table~\ref{tab:formfactors}, we evaluate the decay widths and branching ratios of the semileptonic channels 
$\Lambda_c^+ \to \Lambda^0 e^+ \nu_e$ and 
$\Lambda_c^+ \to \Lambda^0 \mu^+ \nu_\mu$ using Eq.~\eqref{eq10}. The branching fractions derived after the phase-space integrations for the two DA sets are given in Table~\ref{tab:BRresults}.
\begin{table}[ht]
\centering
\caption{Predicted branching fractions (\%) for $\Lambda_c^+ \to \Lambda^0 \ell^+ \nu_\ell$ decays
obtained using two different sets of $\Lambda_c$ distribution amplitudes (DAs).}
\label{tab:BRresults}
\begin{tabular}{lcc}
\toprule
 & $\mathcal{B}(\Lambda_c^+ \to \Lambda^0 e^+ \nu_e)$ 
 & $\mathcal{B}(\Lambda_c^+ \to \Lambda^0 \mu^+ \nu_\mu)$ \\
\midrule
Set I  & $3.56 \pm 0.85 \%$ & $3.51 \pm 0.84$ \\
Set II & $3.90 \pm 0.94 \%$ & $3.88 \pm 0.93$ \\
\bottomrule
\end{tabular}
\end{table}

A comparison with the BESIII measurement indicates that the predictions are in good agreement with experiment.

As mentioned earlier, the $\Lambda_c \to \Lambda \ell \nu_\ell$ decays have
been studied extensively within a variety of theoretical frameworks.
For completeness, in Table~\ref{tab:comparison} we summarize our branching
ratio predictions together with those obtained from various theoretical
approaches and experiments.

\begin{table}[h!]
\centering
\caption{Branching fractions (\%) for $\Lambda_c^+ \to \Lambda^0 \ell^+ \nu_\ell$ 
from various theoretical approaches and experiment.}
\label{tab:comparison}
\begin{tabular}{@{}lcc@{}}
\toprule
Model/Experiment 
&  $\mathcal{B}(\Lambda_c^+ \to \Lambda^0 e^+ \nu_e)$
& $\mathcal{B}(\Lambda_c^+ \to \Lambda^0 \mu^+ \nu_\mu)$  \\
\midrule
LQCD~\cite{Zhang:2025tki}          & $3.88 \pm 0.19$ & $3.75 \pm 0.19$ \\
LQCD~\cite{Meinel:2016dqj}               & $3.80 \pm 0.22$ & $3.69 \pm 0.22$ \\
HBM~\cite{Geng:2022fsr}             & $3.78 \pm 0.25$ & $3.67 \pm 0.23$ \\
CQM~\cite{Gutsche:2015rrt}          & $2.78$          & $2.69$          \\
RQM~\cite{Faustov:2016yza}          & $3.25$          & $3.14$          \\
NRQM~\cite{Hussain:2017lir}         & $3.84$         & $3.72$                   \\ 
QCDSR~\cite{Zhang:2023nxl}          & $3.49 \pm 0.65$ & $3.37 \pm 0.54$ \\
LFQM~\cite{Zhao:2018zcb}                & $1.63$          & \textemdash{}   \\
LFQM~\cite{Geng:2020gjh}              & $3.55 \pm 0.104$& $3.40 \pm 0.102$ \\
LFQM~\cite{Li:2021qod}              & $4.04 \pm 0.75$ & $3.90 \pm 0.73$ \\
$SU(3)_F$~\cite{Geng:2019bfz,He:2021qnc} 
                                    & $3.6 \pm 0.4$   & $3.5 \pm 0.4$   \\
\midrule
ARGUS~\cite{ARGUS:1991bvx}      & \textemdash{}   & $2.37 \pm 0.51$ \\
CLEO~\cite{CLEO:1994lyw}        & \textemdash{}   & $2.68 \pm 0.51$ \\
BESIII~\cite{BESIII:2022ysa,BESIII:2016ffj}        & $3.56 \pm 0.13$ & $3.48 \pm 0.20$ \\
\midrule
This work                & $3.56 \pm 0.85 $ (set-I)   & $3.51 \pm 0.84$ (set-I)   \\
               & $3.90 \pm 0.94$ (set-II)   & $3.88 \pm 0.93$ (set-II)   \\
\bottomrule
\end{tabular}
\end{table}

We observe from Table~\ref{tab:comparison} that our predictions  are consistent with most theoretical approaches
and in particular agree well with the BESIII measurements within uncertainties,
with the exception of the CQM~\cite{Gutsche:2015rrt} and LFQM~\cite{Zhao:2018zcb} results. 

Future refinements of this analysis could include the incorporation of $\mathcal{O}(\alpha_s)$ radiative corrections, the inclusion of subleading $1/m_c$ contributions, and a more precise determination of the nonperturbative input parameters that enter the DA definitions.

\begin{figure}[t]
\includegraphics[width=0.30\textwidth]{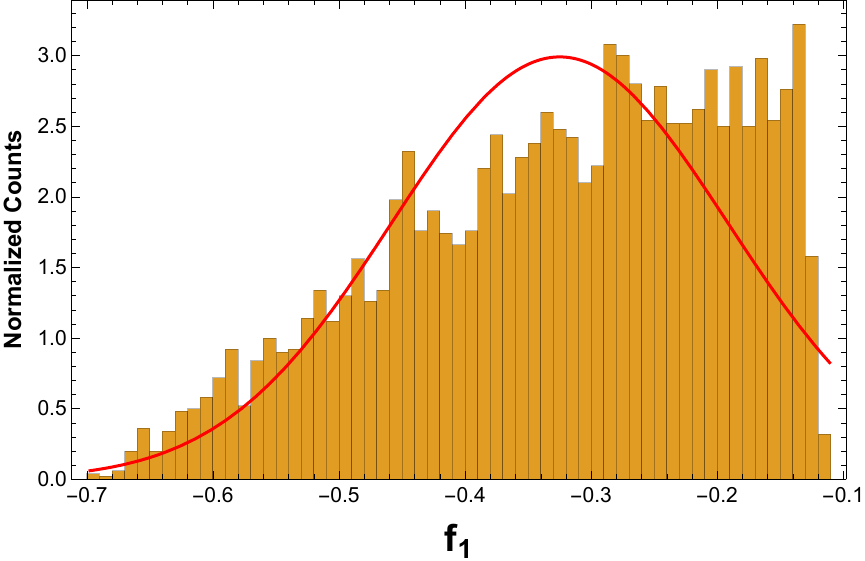}
\includegraphics[width=0.30\textwidth]{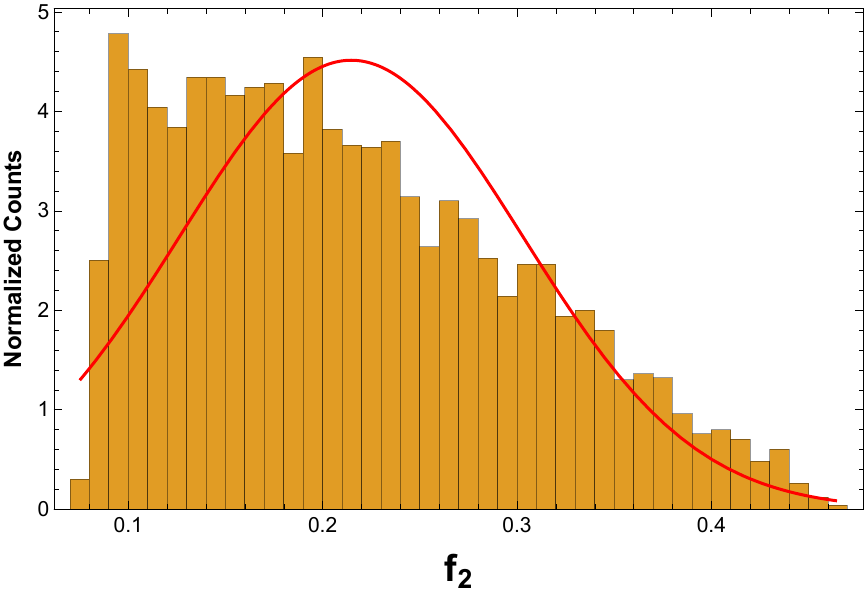}
\includegraphics[width=0.30\textwidth]{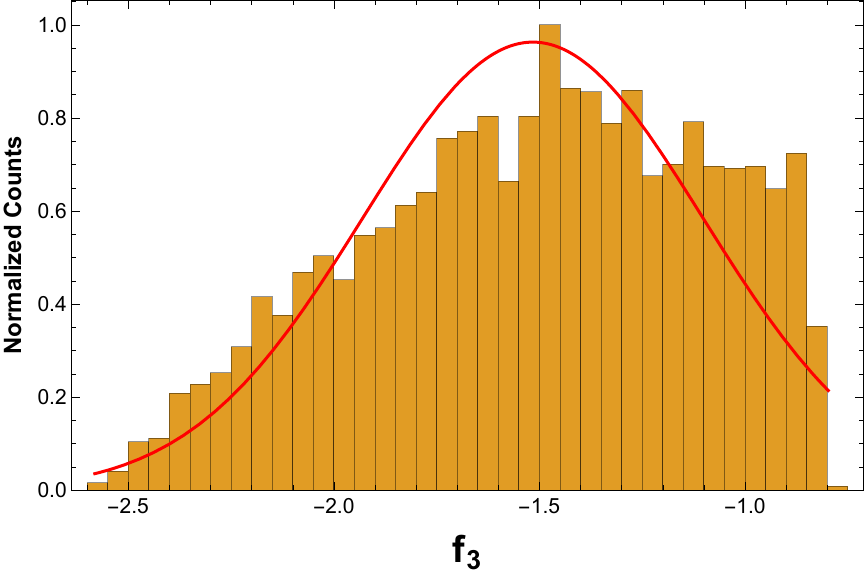} \\
\includegraphics[width=0.30\textwidth]{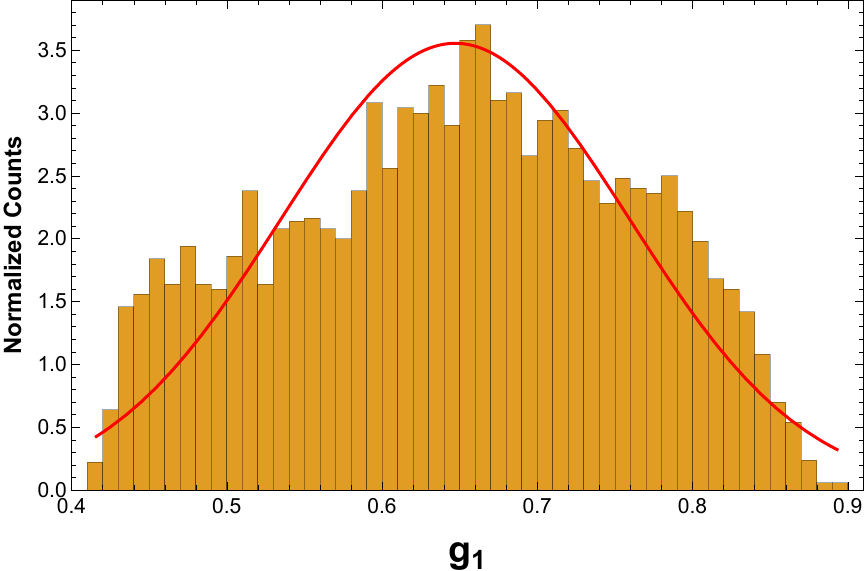}
\includegraphics[width=0.30\textwidth]{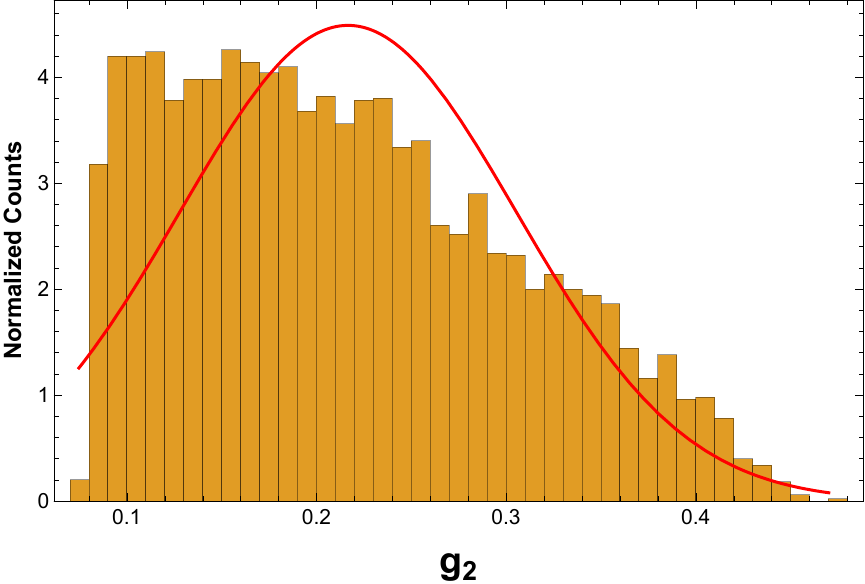}
\includegraphics[width=0.30\textwidth]{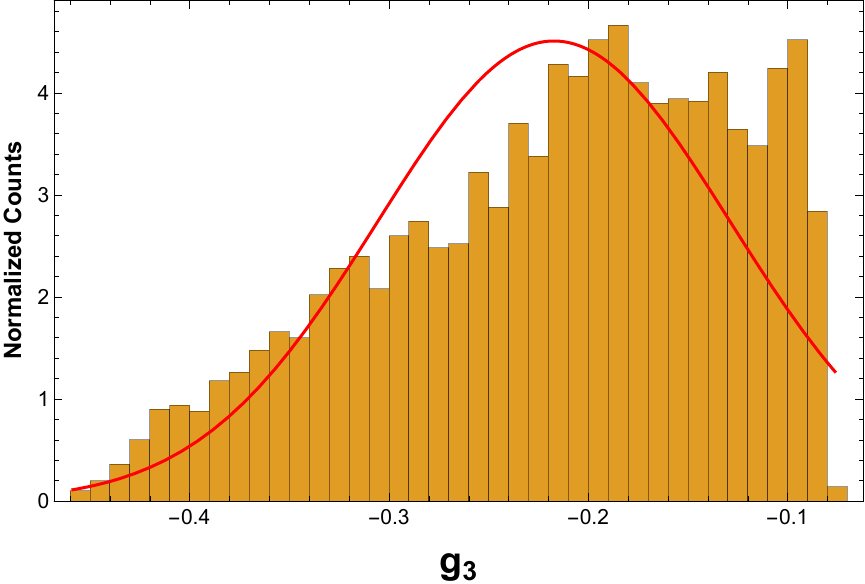}
\caption{Normalized distributions of the $\Lambda_c^+ \to \Lambda^0 \ell^+ \nu$ form factors 
$f_i$ and $g_i$ at $q^2 = 0$ obtained from LCSR. The solid lines represent Gaussian fits 
to the Monte Carlo distributions. For illustration, only the results corresponding to 
the DA set-I are shown.}
\label{fig:setI}
\end{figure}

At the end of this section, we would like to note that we also performed analysis for the other set of Lorentz structures  and found out that the results for the form factors vary slightly.

\section{Conclusion}
\label{sec:conclusion}
Motivated by the recent lattice–experiment tension observed in $\Xi_c \to \Xi \ell^+ \nu_\ell$ transitions, in this work we have investigated the semileptonic decay of the SU(3) partner baryon, $\Lambda_c \to \Lambda \ell^+ \nu_\ell$ ($\ell = e, \mu$), within the framework of light-cone QCD sum rules (LCSR). The sum rules are derived for all form factors, while the numerical analysis is performed using two different sets of $\Lambda_c$ baryon light-cone distribution amplitudes (LCDAs). Using the obtained results for  the form factors, we estimated the branching ratios of the corresponding decays. Our results are in good agreement with the latest BESIII measurements and with lattice-QCD predictions. It is worth noting that our results are also consistent with those obtained from QCD sum rules employing the $\Lambda$-baryon distribution amplitudes~\cite{Zhang:2023nxl}. This agreement indicates that, in contrast to the $\Xi_c$ case, both formulations based on heavy- or light-baryon DAs yield compatible predictions for the $\Lambda_c \to \Lambda$ transition. The absence of any noticeable deviation between theoretical and experimental results in this channel suggests that the discrepancy observed in the $\Xi_c$ decays may not be universal and could arise from either theoretical or experimental sources specific to the $\Xi_c$ sector. Since the $\Lambda_b$ LCDAs were employed as an approximation, subleading $1/m_c$ corrections and the eventual determination of genuine $\Lambda_c$ distribution amplitudes are expected to further improve the precision of the predictions. Future extensions of this analysis may also include $\mathcal{O}(\alpha_s)$ radiative corrections, which would enhance the predictive power of the LCSR framework for charmed baryon decays. Such developments would reduce theoretical uncertainties and provide more robust predictions for comparison with forthcoming lattice and experimental results.


\newpage


\bibliographystyle{utcaps_mod}
\bibliography{all.bib}


\end{document}